\documentstyle[preprint,aps,epsfig]{revtex}

\def\be{\begin{equation}}
\def\ee{\end{equation}}
\def\ep{\epsilon}
\def\t{\tilde}
\def\la{\langle}
\def\ra{\rangle}

\begin{document}
\draft
\title{Spectral Representation Theory for Dielectric Behavior \\ 
       of Nonspherical Cell Suspensions}
\author{J. P. Huang$^1$, K. W. Yu$^1$, Jun Lei$^2$ and Hong Sun$^{2,3}$}
\address{$^1$Department of Physics, The Chinese University of Hong Kong,
 Shatin, NT, Hong Kong}
\address{$^2$Department of Applied Physics, Shanghai Jiao Tong University,
 Shanghai 200 030, China} 
\address{$^3$Department of Physics, University of California, Berkeley,
 California 94720-7300}
\maketitle

\begin{abstract}
Recent experiments revealed that the dielectric dispersion spectrum of 
fission yeast cells in a suspension was mainly composed of two 
sub-dispersions. The low-frequency sub-dispersion depended on the 
cell length, while the high-frequency one was independent of it. 
The cell shape effect was simulated by an ellipsoidal cell model but the 
comparison between theory and experiment was far from being satisfactory. 
Prompted by the discrepancy, we proposed the use of spectral 
representation to analyze more realistic cell models.
We adopted a shell-spheroidal model to analyze the effects of the cell 
membrane. It is found that the dielectric property of the cell membrane 
has only a minor effect on the dispersion magnitude ratio and the 
characteristic frequency ratio.
We further included the effect of rotation of dipole induced by an external 
electric field, and solved the dipole-rotation spheroidal model in the 
spectral representation. Good agreement between theory and experiment 
has been obtained. 
\end{abstract}
\vskip 5mm 

\section{Introduction}

The polarization of biological cells has a wide scope of practical 
applications like manipulation, trapping or separation of biological cells 
\cite{Gimsa,Gimsa99}, and thus an accurate characterization of the 
polarization is needed. 
While the polarization of biological cells can be investigated by the method
of dielectric spectroscopy \cite{Asami80} as well as by the measurement 
of field-induced cell movements \cite{Fuhr,Gimsa91}, 
the former method has a much higher resolution \cite{Gimsa}. 
For biological cells, the main contribution to the dielectric dispersion
is the structural (Maxwell-Wagner) polarization effects \cite{Gimsa}. 
Because of the practical applications, there is a strong need for intuitive 
models as well as simplified equations which describe the parameter 
dependence of the polarization. Thus, various cell models have been
proposed for the analysis of the polarization mechanisms. 
However, due to the complexity of existing theories, these methods have 
not yet found broader acceptance. 

In this work, we propose the use of the spectral representation 
\cite{Bergman} for analyzing the cell models. 
The spectral representation is a rigorous mathematical formalism of 
the effective dielectric constant of a two-phase composite material 
\cite{Bergman}. 
It offers the advantage of the separation of material parameters 
(namely the dielectric constant and conductivity) from the cell
structure information, thus simplifying the study. 
From the spectral representation, one can readily derive the dielectric 
dispersion spectrum, with the dispersion strength as well as the 
characteristic frequency being explicitly expressed in terms of the 
structure parameters and the materials parameters of
the cell suspension (see section II.B below). 
The actual shape of the real and imaginary parts of the permittivity
over the relaxation region can be uniquely determined by the Debye relaxation
spectrum, parametrized by the characteristic frequencies and the dispersion
strengths. So, we can study the impact of these parameters on the dispersion
spectrum directly.

The plan of the paper is organized as follows. 
In the next section, we will review the spectral representation theory 
\cite{Bergman} 
and show that the dielectric dispersion spectrum of a cell suspension 
can be expressed in terms of the spectral representation. 
In section III, we will apply the spectral representation to the various 
cell models and present an alternative approach. 
We show that a better agreement with the experimental data can be achieved.
In section IV, we discuss the effects of dipole rotation. We will show that
the dipole rotation effect has a strong impact on the dispersion spectrum
when the cells are sufficiently long.
Discussion on further applications of our theory will be given.

\section{FORMALISM}

We regard a cell suspension as a composite system consisting of 
biological cells of complex dielectric constant $\t{\ep}_1$ dispersed 
in a host medium of $\t{\ep}_2$. A uniform electric field 
${\bf E}_0=E_0 \hat{\bf z}$ is applied along the $z$-axis. 
We briefly review the spectral representation theory of the effective 
dielectric constant to establish notations.

 \subsection{Spectral representation}
  
The spectral representation is initiated by solving the differential equation
 \be
\nabla\cdot\left[\left(1-\frac{1}{\t{s}}\eta({\bf r})\right)
 \nabla \phi({\bf r})\right]=0,
 \ee
where $\t{s}=\t{\ep}_2/(\t{\ep}_2-\t{\ep}_1)$ denotes the relevant 
material parameter and $\eta({\bf r})$ is the characteristic function of 
the cell structure. The electric potential $\phi ({\bf r})$ can be 
solved formally \cite{Bergman}
 \be
\phi({\bf r})=-E_0 z+\frac{1}{\t{s}}\int d{\bf r}'\eta({\bf r}')
 \nabla 'G_0({\bf r}-{\bf r}')\cdot \nabla '\phi ({\bf r}'),
 \ee
where $G_0({\bf r}-{\bf r}')=1/4\pi|{\bf r}-{\bf r}'|$ is the free space 
Green's function. By denoting an operator
 \be
\Gamma=\int d{\bf r}'\eta({\bf r}')\nabla 'G_0({\bf r}-{\bf r}')\cdot\nabla ',
 \ee
and the corresponding inner product
 \be
\la \phi|\psi\ra=\int d{\bf r}\eta ({\bf r})
 \nabla\phi^*\cdot\nabla\psi,
 \ee
it is easy to show that $\Gamma$ is a Hermitean operator. 
Let $s_n$ and $|n\ra$ be the eigenvalue and eigenfunction of $\Gamma$
such that $\Gamma |n\ra = s_n |n\ra$,
where $0\le s_n<1$ is a real eigenvalue.
The integral equation can be solved symbolically:
 \be
|\phi\ra = -{\t{s} \over \t{s}-\Gamma} |z\ra E_0.
 \ee
From the solution, we obtain the electric field and hence compute the 
effective dielectric constant $\t{\ep}_e$ in the spectral representation.
We further define the reduced effective dielectric function \cite{Bergman}:
 \be
F(\t{s})=1-{\t{\ep}_e \over \t{\ep}_2} = -{1\over \t{s}VE_0} 
 \la z|\phi\ra .
 \ee
By inserting the complete set $1 = \sum_n |n\ra\la n|$, we find
 \be
F(\t{s})={1\over V} \sum_n {\la z|n\ra \la n|z\ra \over 
 \t{s}-s_n}
 =\sum_n {F_n \over \t{s}-s_n}.
 \label{Fs}
 \ee
$F_n$ is defined as the spectral function:
 \be
F_n={1\over V} \la z|n\ra \la n|z\ra,
 \ee
which satisfies a sum rule \cite{Bergman}:
 \be
\sum_n F_n={1\over V} \sum_n \la z|n\ra \la n|z\ra
= {1\over V} \la z|z\ra = V_1/V=p,
 \ee
where $V_1$ is the total volume of the suspending cells and $p$ 
the volume fraction of the cells.

 \subsection{Dielectric dispersion spectrum}
 
For cells of arbitrary shape, the eigenvalue problem of the $\Gamma$ 
operator can only be solved numerically. 
However, analytic solutions can be obtained
for isolated spherical and ellipsoidal cells.
For dilute suspensions of prolate spheroidal cells, the cells can be 
regarded as noninteracting.
The problem is simplified to the calculation of $s_n$ and $|n\ra$ 
with a single cell, which can be solved exactly.
Only two of the $F_n$ are nonzero, due to the orthogonality of 
$|n\ra$ with $|z\ra$.

Thus, in subsequent studies, we restrict ourselves to two poles ($n=1,2$). 
From Eq.(\ref{Fs}), the effective dielectric constant is written in the 
spectral representation:
 \be
\t{\ep}_e=\t{\ep}_2\left(1-\sum_{n=1}^2\frac{F_n}{\t{s}-s_n}\right).
\label{sp}
 \ee
After substituting $\t{\ep}_1=\ep_1+\sigma_1/j2\pi f$ and 
$\t{\ep}_2=\ep_2+\sigma_2/j2\pi f$ into Eq.(\ref{sp}), where $\ep$
and $\sigma$ are the real and imaginary parts of the complex dielectric 
constant, 
we rewrite the effective dielectric constant
after simple manipulations:
 \be
\t{\ep}_e=\ep_H + \sum_{n=1}^2 \frac{\Delta\ep_n}
 {1+jf/f^{c}_{n}} + \frac{\sigma_L}{j2\pi f}, 
 \ee
where $\ep_H$ and $\sigma_L$ are the high-frequency dielectric
constant and the low-frequency conductivity respectively, while
$\Delta\ep_n$ are the dispersion magnitudes,
$f^{c}_{n}$ are the characteristic frequencies of the $n$th sub-dispersion.
We obtain the dispersion magnitudes $\Delta \ep_n$ and the 
characteristic frequencies $f^c_n$, respectively \cite{Lei}:
\begin{eqnarray}
  \Delta\ep_1&=&F_1\ep_2\frac{s_1(s-t)^2}{s(s-s_1)(t-s_1)^2},\nonumber\\
  \Delta\ep_2&=&F_2\ep_2\frac{s_2(s-t)^2}{s(s-s_2)(t-s_2)^2},\nonumber\\
  f_1^c&=&\frac{\sigma_2s(t-s_1)}{2\pi\ep_2t(s-s_1)},\nonumber\\
  f_2^c&=&\frac{\sigma_2s(t-s_2)}{2\pi\ep_2t(s-s_2)},\nonumber
\end{eqnarray}
where $s=\ep_2/(\ep_2-\ep_1)$ and $t=\sigma_2/(\sigma_2-\sigma_1)$. 
To compare with experiment data \cite{Asami},
we express the dispersion magnitude ratio and characteristic frequency 
ratio as
\begin{eqnarray}
\frac{\Delta\ep_1}{\Delta\ep_2}&=&\frac{F_1}{F_2}\cdot 
\frac{s_1(s-s_2)(t-s_2)^2}{s_2(s-s_1)(t-s_1)^2}, \label{mag}\\
\frac{f_2^c}{f_1^c}&=&\frac{(t-s_2)(s-s_1)}{(t-s_1)(s-s_2)}. \label{freq}
\end{eqnarray}

\section{Applications to Various Cell Models}

In a recent work \cite{Lei}, we adopted the spheroidal model 
(SM) to analyze the cell suspensions.
Here we briefly review the analytic results of the model: 
$$
s_1=L_z,\ \  s_2=L_{xy},\ \
F_1=\frac{1}{3}p,\ \ F_2=\frac{2}{3}p.
$$
where
\begin{eqnarray}
L_z &=& -\frac{1}{q^2-1}+\frac{q}{(q^2-1)^{3/2}}\ln (q+\sqrt{q^2-1}), 
 \nonumber\\ 
L_{xy}&=&(1-L_z)/2
 \nonumber
\end{eqnarray}
are the depolarization factors along the $z$-axis and $x$-(or $y$-) axis 
of the prolate spheroid and $q$ is the ratio of length $L$ to diameter $D$.

In the spheroidal model, we neglected the presence of a cell membrane. 
To study this effect, we put forward the shell-spheroidal model (SSM) here.
In this case, the biological cells are modelled as shell-spheroidal ones
with a spheroidal core of complex dielectric constant $\ep_1$, 
covered with a confocal spheroidal shell of $\ep_s$.
For a small volume fraction $p$ of shelled spheroidal cells embedded in a
host medium of complex dielectric constant $\ep_2$, 
the effective dielectric constant $\t{\ep}_e$ is given by the 
dilute-limit expression:
$$
\t{\ep}_e=\t{\ep}_2+p\t{\ep}_2(b_z+2b_{xy}).
$$
where $b_z$ is the dipole factor for a single-shelled spheroidal cell 
along the $z$-axis \cite{Gao}:
$$b_z=\frac{1}{3}\frac{(\t{\ep}_s-\t{\ep}_2)
 [\t{\ep}_s+L_z(\t{\ep}_1-\t{\ep}_s)]
 +(\t{\ep}_1-\t{\ep}_s)y[\t{\ep}_s+L_z(\t{\ep}_2-\t{\ep}_s)]}
 {(\t{\ep}_s-\t{\ep}_1)(\t{\ep}_2-\t{\ep}_s)yL_z(1-L_z)
 +[\t{\ep}_s+L_z(\t{\ep}_1-\t{\ep}_s)]
 [\t{\ep}_2+L_z(\t{\ep}_s-\t{\ep}_2)]},
$$
where $y$ is the volume ratio of core to the whole shelled spheroid,
while $b_{xy}$ indicates the dipole factor along the $x$- (or $y$-) axis,
which can be obtained by replacing the subscript $z$ with $xy$ in the 
expression of $b_z$.
As a matter of fact, the cell suspension consisting of shell-spheroidal 
cells dispersed in a host medium is a three-phase system. 
Although the spectral representation was generally valid for two-phase 
composites, we have recently shown that it applies to composites of coated 
spheres as well as to coated spherical particles randomly embedded in a 
host medium \cite{Yuen97}. 
Note the sum rule $\sum F_n=p$ is no longer valid.
Similarly, one can show that the spectral representaton also applies to 
the present system consisting of spheroidal cells with shells of complex 
dielectric constant $\t{\ep}_s$ dispersed in a host. 
The effective dielectric constant is then given by 
 \be
\t{\ep}_e=\t{\ep}_2
 \left[1-\left(\sum_{n=1}^2\frac{F_n}{\t{s}-s_n}+N.P.\right)\right]
\ee
with $N.P.$ being the nonresonant part which vanishes in the limit of 
unshelled spheroidal inclusions, where 
\begin{eqnarray}
s_1&=&\frac{L_z[1+(x-1)y+L_z(-1+x+y-xy)]}
 {x-L_z(x-1)^2(y-1)+L_z^2(x-1)^2(y-1)},\nonumber\\
s_2&=&\frac{L_{xy}[1+(x-1)y+L_{xy}(-1+x+y-xy)]}
 {x-L_{xy}(x-1)^2(y-1)+L_{xy}^2(x-1)^2(y-1)},\nonumber\\
F_1&=&\frac{p x^2 y}{3[x-L_z(x-1)^2(y-1)+L_z^2(x-1)^2(y-1)]^2},
 \nonumber\\
F_2&=&\frac{2 p x^2 y}{3[x-L_{xy}(x-1)^2(y-1)+L_{xy}^2(x-1)^2(y-1)]^2},
 \nonumber
\end{eqnarray}
where $x=\t{\ep}_s/\t{\ep}_2$.
We omit the complicated expression for the nonresonant part here.
In what follows, for the sake of convenience, we assume: 
(1) $y$ is a constant for all coated spheroid; (2) $x$ is a real number.

In Fig.1, we plot the structure parameters $F_n$ and $s_n$ versus $x$ for 
various $y$ and for (a) $q=3.46$, (b) $7.17$ and (c) $10.24$, respectively. 
In all case, $p=0.01$. We find $F_n$ is strongly dependent on $y$ for 
$x>0.5$, whereas it is not the case for $s_n$. It may be concluded 
that the dielectric property of the cell membrane has a minor effect on 
the dispersion magnitude ratio, 
but plays no role in the characteristic frequency ratio.

To investigate the validity of these models, we compare to experimental 
data, which was extracted by using a temperature sensitive cell division 
cycle mutant of fission yeast, $cdc25$-$22$ \cite{Asami}.
Asami's theory \cite{Asami} results are also plotted for comparison. 
From Fig.2, it is evident that our model gives a better comparison with 
experimental data than Asami's theory. The reason for the improvement 
lies in the introduction of the conductivity contrast $t$ by using of the 
spectral representation. As stated in Ref.\cite{Lei}, the large difference 
between our model and Asami's theory is due to a large 
$\sigma_1 \gg \sigma_2$ used in contrast to Asami's claim
$\sigma_1\approx \sigma_2$ \cite{Asami}.  
We further find that SSM provides a better fit than SM for the dispersion
magnitude ratio $\Delta \ep_1/\Delta \ep_2$, while SSM yields the same 
results as those of SM for the frequency ratio $f_2^c/f_1^c$ (both curves
overlap in the right panel of Fig.2), indicating that the dielectric 
property of a cell membrane is indeed unimportant.

\section{Effects of dipole rotation}

According to the numerical results, we find that the SSM provides a better 
fitting with previous experimental data than SM, 
but this improvement is actually too small.
In other words, the dielectric properties of a cell membrane does not play 
an important role in dielectric dispersion spectrum.
But, those numerical result will also show that both SM and SSM cannot 
obtain a good agreement with experimental data.
In the presence of an electric field, cells of large length may rotate
in favor of the applied field, thus we propose another model, namely 
the dipole-rotation spheroidal model (DRSM) to obtain a better fitting.
 
When the cells are long enough, the rotation of dipole becomes very 
important with the external electric field under consideration, 
and the system is in general anisotropic.
We have to take into account the effect of dipole rotation on $F_1$ and 
$F_2$, even for a weak electric field. Let us compute them from a 
thermodynamic consideration.
We will show that they in general depends on $q$.
  
Consider a spheroidal cell in an electric field $E_0$. Its long axis 
makes an angle $\theta$ with the field. The dipole energy of the cell is
 \be
E_d(q,\theta)=-{\rm Re} \left[\frac{\t{\ep}_2D^3E_0^2}{16}
 q(b_z\cos^2\theta +b_{xy}\sin^2\theta)\right],
 \label{dipole}
 \ee
where $b_z$ and $b_{xy}$ are dipole factors along and perpendicular to 
the long axis: 
$$
b_z=\frac{1}{3(L_z-\t{s})},\  \ b_{xy}=\frac{1}{3(L_{xy}-\t{s})}.
$$
Eq.(\ref{dipole}) can be understood by the energy approach. 
For simplicity, suppose the major axes of the cells all lie along the 
electric field, i.e., $\theta=0$, then the induced dipole moments of 
the cells give a contribution to the effective dielectric constant. 
In the dilute limit,
$$
\t{\ep}_e=\t{\ep}_2+3 p \t{\ep}_2 b_z,
$$
where $p=V_1/V$ is the volume fraction of the cells.
For a fixed external field condition, the total electrostatic energy density 
of the suspension is given by $E_t=-{\rm Re} (\t{\ep}_e E_0^2/8\pi)$, 
which is equal to $-{\rm Re} (\t{\ep}_2 E_0^2/8\pi) + E_d/V$, 
and hence the desired results. 

We showed that the conductivity contrast $t$ attains a small negative value 
\cite{Lei} and thus the complex material parameter $\t{s}$ can be 
approximated by $t$. 
Consequently, both $b_z$ and $b_{xy}$ have positive values. 
The probability is given by the Boltzmann factor
 \be
\rho(q,\theta)=A e^{-E_d(q,\theta)/k_BT}
 \ee
where $A$ is a normalization factor such that 
$\int\rho(q,\theta) {\rm d}\Omega=1$, where $\Omega$ is the solid angle, 
$d\Omega=\sin\theta d\theta d\varphi$.
We can calculate $F_1$ and $F_2$ by the following integrals 
\begin{eqnarray}
F_1(q)=p\int\rho(q,\theta)\cos^2\theta {\rm d}\Omega,\ \ \
F_2(q)=p\int\rho(q,\theta)\sin^2\theta {\rm d}\Omega.
\end{eqnarray}
The $F_1(q)/F_2(q)$ ratio may be obtained by integrating with respect to 
$\theta$ from $0$ to $\pi/2$ by symmetry. 
In the absence of an electric field, $E_d(q,\theta)=0$ and 
$\rho(q,\theta)$ equals to a uniform distribution. 
In which case, we obtain $F_1=p/3$ and $F_2=2p/3$, and hence $F_1/F_2=0.5$. 
When the electric field is weak enough,
the ratio is still constant and $F_1/F_2=0.5$. 
Otherwise, the ratio will increase rapidly with $q$. 
For $q=1$, $b_z=b_{xy}$ and $F_1/F_2=0.5$ always.
The above result implies that both $F_1$ and $F_2$ depend strongly on 
$q$ when there is an electric field.
For large $q$, $b_z\gg b_{xy}$, the spheroids tend to align with the 
applied field and hence $F_1/F_2$ becomes very large.  
 
It is found that the mean cell length depends on the cultivation time, 
whereas the diameter is almost unchanged in an experiment \cite{Asami}, 
which will be applied to compare the different models.
In the following numerical calculation, without loss of generality, 
we neglect the small imaginary part of $\t{\ep}_2$, and define 
a new parameter $\xi$:
 \be
\xi=\frac{\ep_2D^3E_0^2}{16 k_{B}T}
 \ee
which characterizes the electric field strength. 

We can readily obtain the dispersion magnitude ratio 
$\Delta \ep_1/\Delta \ep_2$ as well as the dispersion frequency ratio 
$f_2^c/f_1^c$ by substituting the results of $F_1(q)/F_2(q)$ 
into Eqs.(\ref{mag}) and (\ref{freq}), 
and setting $s_1=L_z$ and $s_2=L_{xy}$.
In Fig.3, $F_1/F_2$ is plotted versus $q$. It is shown that $F_1/F_2$ 
depends strongly on the axial ratio $q$, especially 
for large $q$ or strong magnitude of external electric field.

To compare with experimental data in Fig.2, we obtain good agreement 
in the DRSM with $\xi=0.017$, which corresponds to a weak field
$E_0 \approx 0.1$ V/m. The results show that dipole rotation indeed plays 
an important role in the dielectric dispersion --
we cannot neglect the effect of the rotation of dipole induced by the 
applied electric field, especially when the average length of cell is large.
In addition, good agreement exists only for large cytoplasmic conductivity, 
as attributed to a higher ion concentration in their cytoplasm to avoid 
the shrinkage of cells due to a loss of water across the cell membrane.
  
\section{DISCUSSION AND CONCLUSION}
 
Here we would like to make a few comments.
At low frequencies, the cell membrane effectively insulates the interior 
of cell. In other words, a potential builds up entirely over the cell 
membrane, leaving the interior of cell rather inactive to the field 
\cite{Gimsa}.
Thus $\ep_1 \ll \ep_2$ and $s=1^+$ and it is reasonable to use $s=1.001$ 
as fitting parameter. On the other hand, we assume that the host medium
has low loss and $\sigma_2 \approx 0$ and at the same time a large
cytoplasmic conductivity $\sigma_1 \gg \sigma_2$. 
Thus $t=0^-$ and it is reasonable to use $t=-0.0001$ to fit the data.

The resulting equations [Eqs.(\ref{mag}) and (\ref{freq})] are indeed 
simple equations arising from the spectral representation. 
These equations serve as a basis which describe the parameter dependence 
of the polarization and thereby enhances the applicability of various 
cell models for the analysis of the polarization mechanisms. 
In this connection, the shell-spheroid model may readily be extended to 
multi-shell cell model.
However, we believe that the multi-shell nature of the cell may have 
a minor effect on the dispersion magnitude ratio as well as on the 
characteristic frequency ratio.

In the presence of external electric fields, field-induced motions such as
rotation of cells, dielectrophoretic motion or vibrational motion
may have a significant impact on the dielectric dispersion spectrum. 
With the recent advent of experimental techniques such as automated video 
analysis \cite{Gasperis} as well as light scattering methods\cite{Gimsa99}, 
the cell movements can be accurately monitored.
For purely rotational motions, the distribution of surface charge on the
cell surfaces may deviate significantly from the equilibrium distribution 
for cells at rest, leading to a change in the polarization relaxation 
and in the dielectric dispersion spectrum. 
In this regard, our recent work on dynamic electrorheological effects 
\cite{Wan}, in which the suspended particles can have rotational motions, 
may be applied to cell rotational motions. 
In Ref.\cite{Wan}, we found that the particles' rotational motions do 
change the polarization relaxation substantially.

In this work, we considered a monodisperse cell suspension, 
in which the cells are of the same shape (i.e., same length and diameter). 
While the diameter of the cells may remain constant during the cultivation 
process, the cells may possess a wide distribution of cell lengths 
\cite{Asami}. 
A modified theory, which takes the distribution of length into account, 
is urgently needed and our spectral representation theory will certainly 
help. 
In this connection, we may apply a strong dc electric field (in addition to 
the ac probe field) to help separating the long cells from the short ones. 
Our results indicate that even in a moderate field, 
the long cells can easily be aligned with the applied field, 
while the short ones remain essentially randomly oriented. 
In this way, an emphasis of dispersion spectrum of the long cells can be
made possible.

In summary, prompted by the discrepancy between recent theory and experiment
on fission yeast cells, we have proposed the use of spectral representation 
to analyze more realistic cell models.
We adopted a shell-spheroidal model to analyze the effects of the cell 
membrane. It is found that the presence of a membrane has only a minor 
effect on the dispersion ratio, but plays no role in the frequency ratio.
We further included the effect of rotation of dipole induced by an external 
electric field.
It has been found that the dipole-rotation effect plays an important role 
in the dispersion magnitude, but it does not change the characteristic 
frequency ratio. We obtained good agreement between theory and experiment 
when dipole-rotation effect is included.

\section*{Acknowledgments}
This work was supported by the Research Grants Council of the Hong Kong 
SAR Government. 
J. P. H. is grateful to Dr. L. Gao and Dr. C. Xu for fruitful discussion.
K. W. Y. acknowledges useful discussion with Prof. G. Q. Gu.

\begin{figure}[h]
\caption{For SSM, $F_1$, $F_2$, $s_1$ and $s_2$ are plotted versus the 
 dielectric constant ratio $x$ for different thickness parameter $y$: 
 (a) $q=3.46$; (b) $q=7.17$; (c) $q=10.24$.}
\end{figure}

\begin{figure}[h]
\caption{Ratios of the dispersion magnitudes and the characteristic 
frequencies are plotted versus $q$. 
Asami's theory: $\sigma_1 \approx \sigma_2$;
SM: $t=-0.0014$, $s=5.0$; 
SSM: $t=-0.0014$, $s=5.0$, $x=2$, $y=0.8$; 
DRSM: $t=-0.0001$, $s=1.001$, $\xi=0.017$ 
(i.e., $E_0$ is about $0.1$V/m).
Note that the curves of SM and SSM overlap in the right panel,
while they are quite close in the left panel.}
\end{figure}

\begin{figure}[h]
\caption{For DRSM, the ratio $F_1/F_2$ is plotted versus $q$ for 
 different electric field strength parameter $\xi$.}
\end{figure}

\newpage
\centerline{\epsfig{file=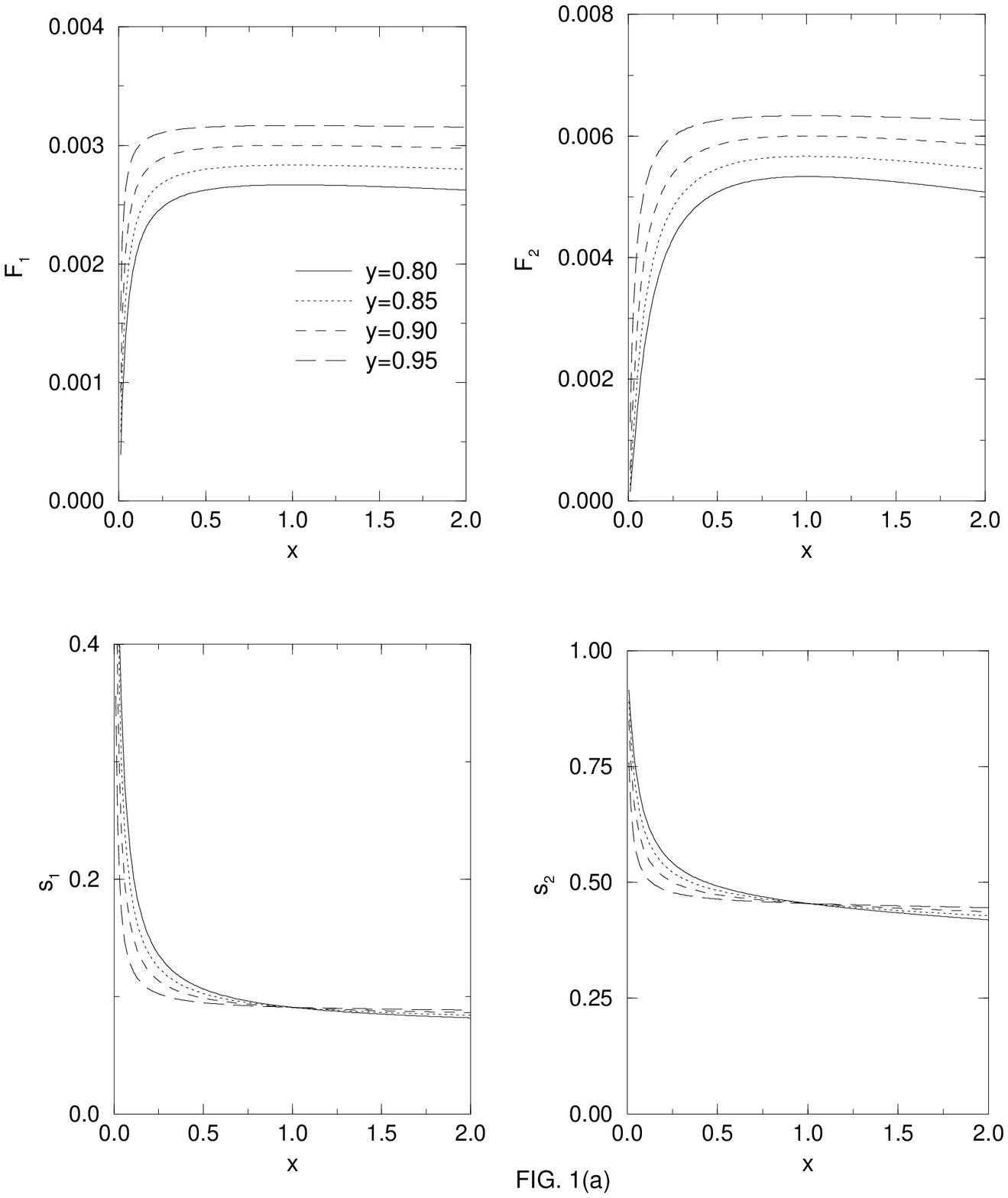,width=\linewidth}}
\centerline{Fig.1(a)/Huang, Yu, Lei, Sun}
\centerline{\epsfig{file=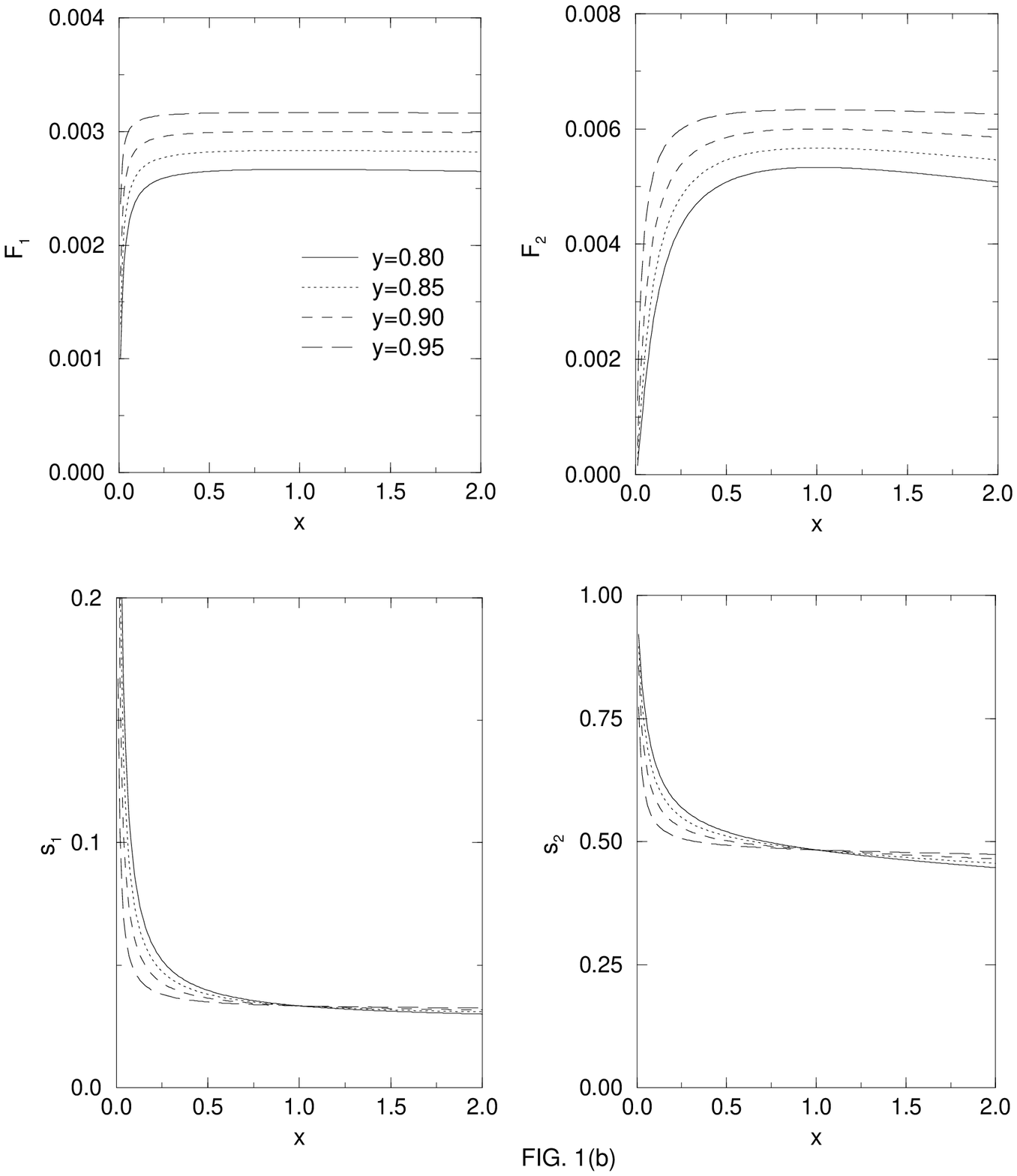,width=\linewidth}}
\centerline{Fig.1(b)/Huang, Yu, Lei, Sun}
\centerline{\epsfig{file=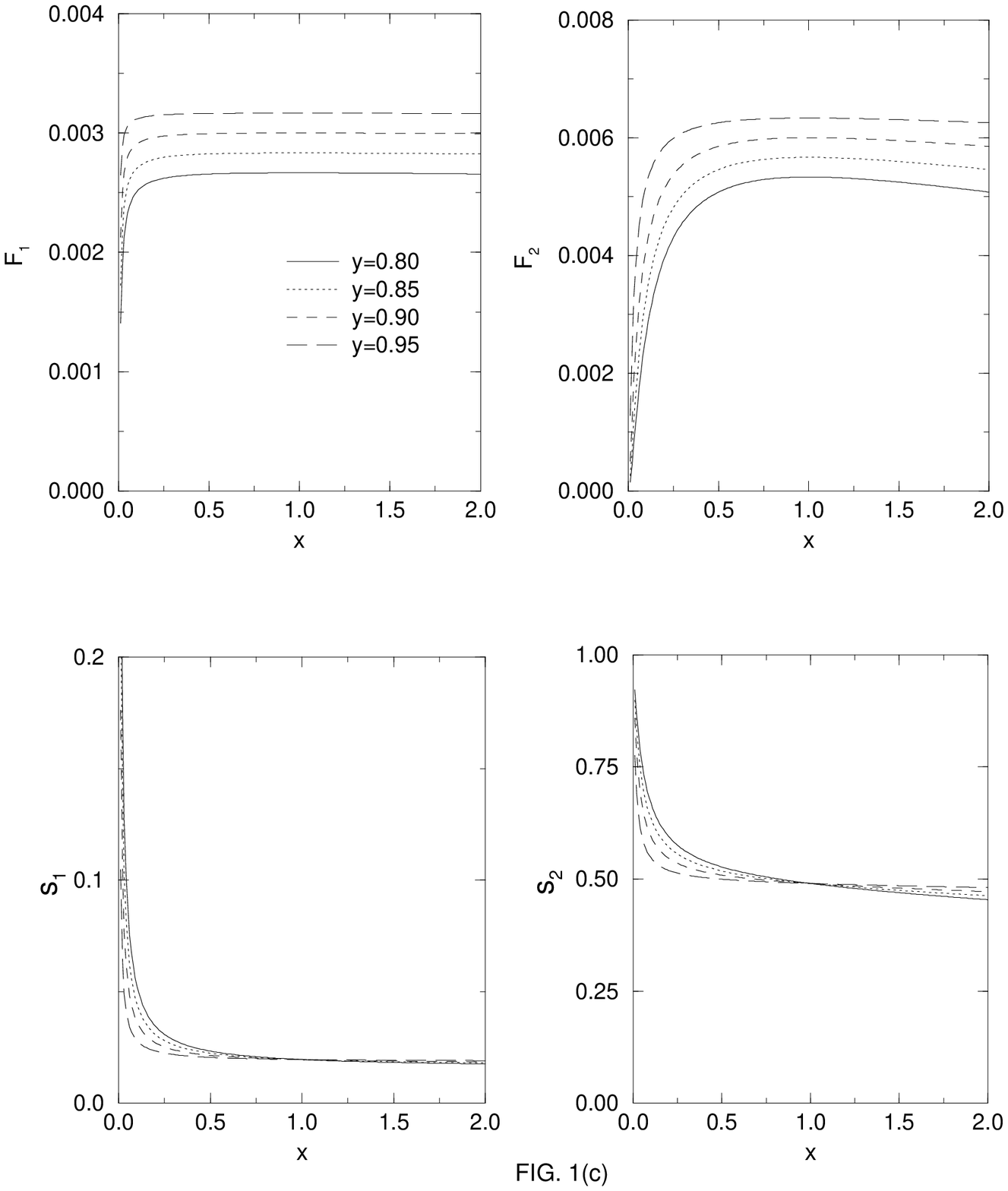,width=\linewidth}}
\centerline{Fig.1(c)/Huang, Yu, Lei, Sun}

\centerline{\epsfig{file=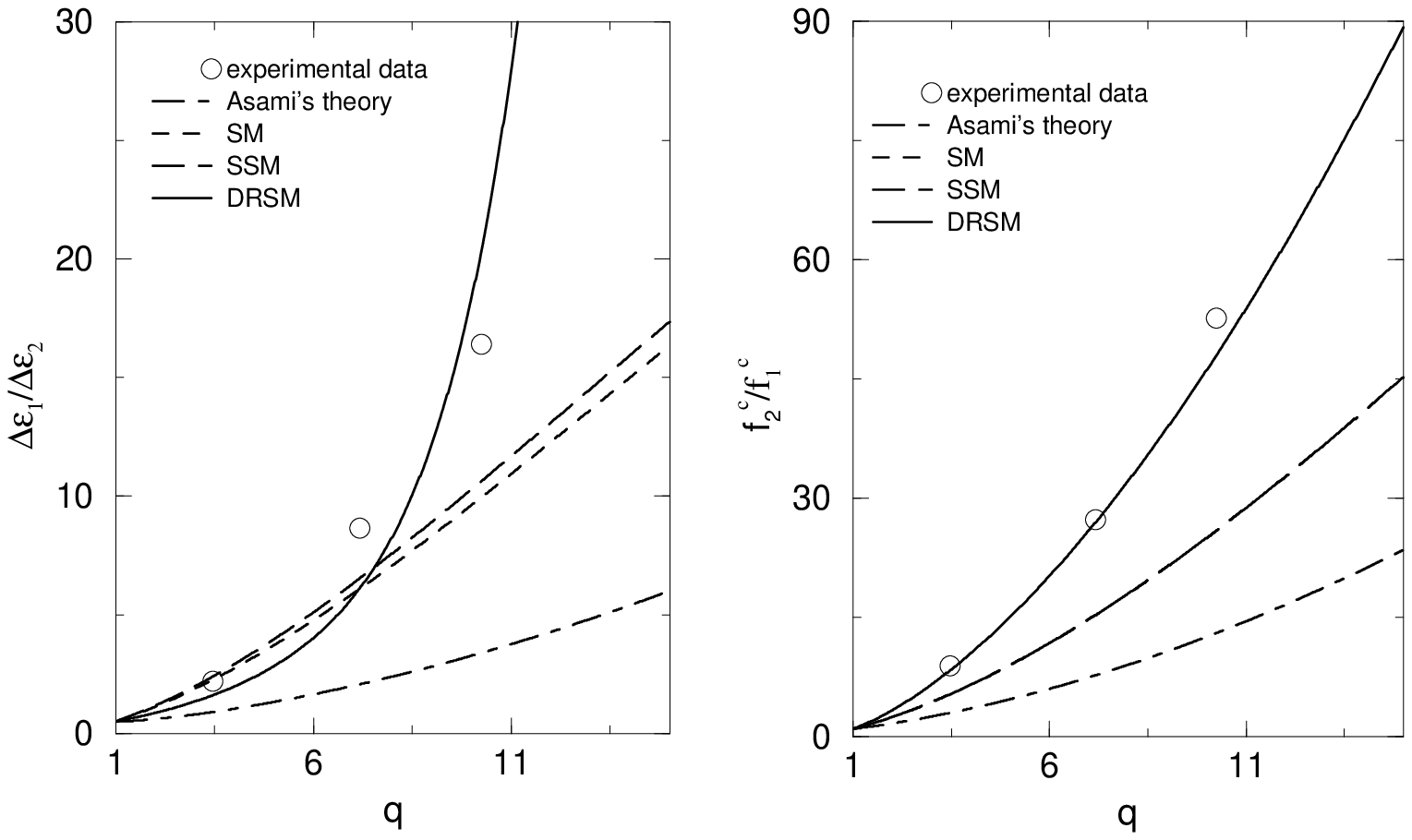,width=\linewidth}}
\centerline{Fig.2/Huang, Yu, Lei, Sun}

\centerline{\epsfig{file=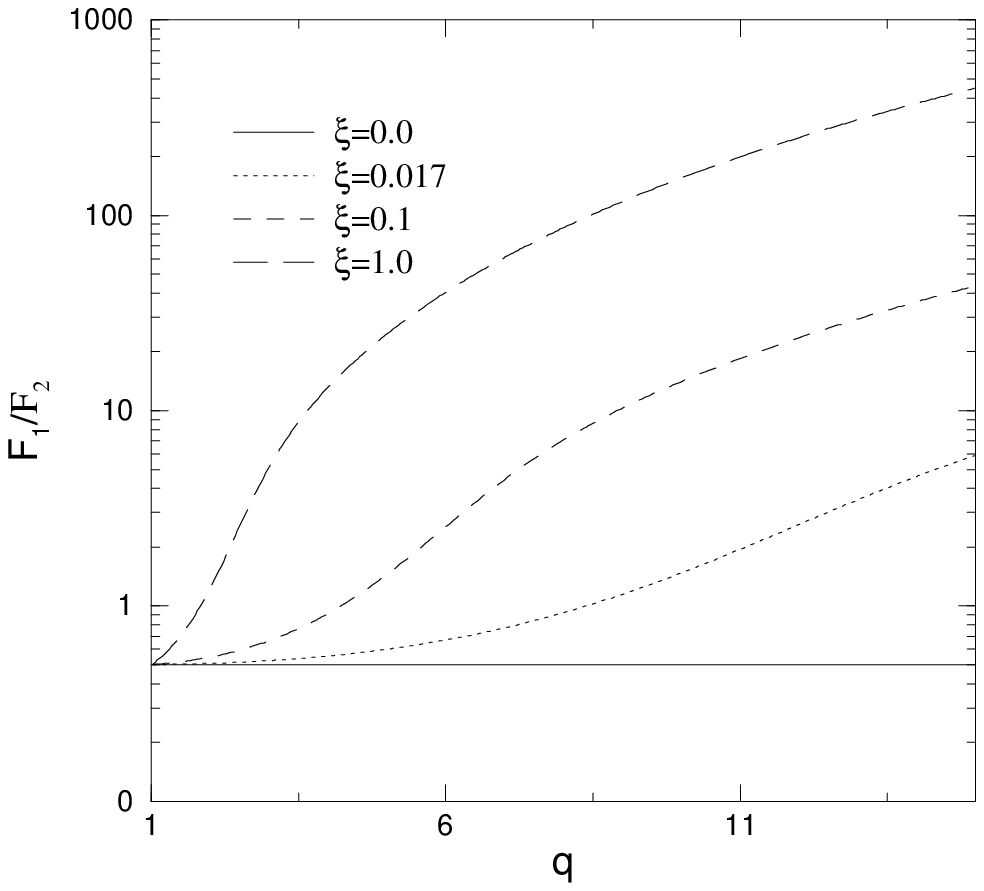,width=\linewidth}}
\centerline{Fig.3/Huang, Yu, Lei, Sun}

\end{document}